# Una mirada alternativa a la producción científica de la Sociología española ¿Qué nos dicen las altmetrics?

Daniel Torres-Salinas[1,2], Wenceslao Arroyo-Machado[1], Nicolás Robinson-García[3]

1. Universidad de Granada, Dpt. de Información y Comunicación y Medialab UGR
2. Universidad de Granada, EC3metrics
3. TU Delft, Delft Institute of Applied Mathematics

**Resumen**: En los últimos años se han introducido nuevos indicadores denominados altmétricas con los que medir el impacto de la actividad científica. Estos indicadores se obtienen a través de las menciones realizadas desde diferentes medios sociales, existiendo varios agregadores de estos datos que reunen varios de ellos en una misma base de datos, siendo Altmetric.com el más popular. Sin embargo, a pesar de la popularización de estas métricas se han venido manifestado diversas limitaciones en su uso. Es por ello que el objetivo de este trabajo es doble, (1) por un lado mostrar las posibilidades de las técnicas altmétricas aplicadas a las ciencias sociales española en general y la sociología en particular; (2) analizar críticamente los resultados para observar las limitaciones de éstos indicadores; (3) comprobar si realmente pueden añadir información útil que pueda servir para describir un campo científico y (4) ver las causas que hacen que en estos campos no se pueden aplicar altmetrics.

**Palabras clave**: Sociología, Ciencias Sociales, altmétricas, Altmetric.com

**Abstract**: *In recent years, new indicators known as altmetrics have been introduced to measure the impact of scientific activity. These indicators are obtained through the mentions realised from different social media, existing several aggregators of these data that collect several of them in the same database, being Altmetric.com the most popular. However, in spite of the popularization of these metrics, several limitations in their use have been manifested. For this reason, rhe objective of this work is twofold: (1) to show the possibilities of altmetric techniques applied to the Spanish social sciences in general and sociology in particular; (2) to critically analyse the results to observe the limitations of these indicators; (3) to check whether they can really add useful information that can be used to describe a scientific field and (4) to see the reasons why altmetrics cannot be applied in these fields.*

# 1. Introducción

Las transformaciones sociales experimentadas a lo largo de la última década, enmarcadas en la profunda crisis económica vivida en España y Europa así como la expansión de las tecnologías digitales generando nuevos medios de participación, han renovado la exigencia de transparencia e impacto social en la labor científica desarrollada en universidades y centros de investigación. El discurso político ha asumido también estas exigencias de impacto social dando cabida en la regulación a propuestas y enfoques que recogen nuevas formas de interacción entre ciencia y sociedad (Nowotny, Scott y Gibbons, 2001). El objetivo que pretenden alcanzar con esto es el de alinear los esfuerzos científicos con las demandas y necesidades de la sociedad, articulando una agenda científica común, integrada y transparente. Esto ha llevado a algunos países a desarrollar iniciativas de evaluación del impacto social, siendo pionero el Research Excellence Framework (REF) de 2014 en el Reino Unido (Stern, 2016), donde por primera vez en un ejercicio de evaluación nacional, se valoraba el impacto social de la investigación a través de casos de estudio (Bornmann, Haunschild y Adams, 2019).

Desde el lado de la Cienciometría, y en general desde los Estudios de Ciencia y Tecnología, todo ello implica un cambio de paradigma en los métodos de evaluación que se ha materializado con la introducción de nuevos indicadores que permitan captar la atención social que recibe la investigación. En este capítulo de libro ofrecemos una mirada diferente hacia la producción de la sociología española, alternativa a la presentada en el capítulo de Jiménez-Contreras, Repiso y Ruiz-Pérez en este mismo libro con el objetivo de analizar y entender cuál es la percepción que tiene la sociedad de los trabajos, temas y líneas de investigación sociológicas. Hasta hace unos años, la comunidad científica española ha estado muy centrada en analizar y comprender la recepción y visibilidad que tiene su producción científica, dirigiendo sus esfuerzos hacia la internacionalización de su investigación con el fin de acabar con años de aislamiento científico y alcanzar un nivel productivo de relevancia internacional a la altura del resto de países avanzados (Jiménez-Contreras, Moya Anegón y Delgado López-Cózar, 2003). Pero las nuevas tendencias en evaluación y política científica apuntan hacia un mayor peso de la transferencia de conocimiento e impacto social en la agenda científica (Frodeman y Parker, 2009; Holbrook, 2012; Meljgaard et al., 2018; Wilsdon et al., 2015; Wouters et al., 2019). Esto ha dado lugar a nuevas propuestas de indicadores y metodologías que permitan capturar la atención mediática y la recepción social de los trabajos de investigación.

La propuesta que mayor relevancia ha tenido hasta el momento son los llamados indicadores altmétricos. Nacidas como crítica y alternativa a los indicadores de citación tradicionales (Priem et al., 2010; Torres-Salinas, Cabezas-Clavijo y Jiménez-Contreras, 2013), las altmétricas o *altmetrics* se definieron originariamente como "el estudio y uso de medidas de impacto académico basado en la actividad en herramientas y medios electrónicos en línea"[1] (Priem, 2014, p. 266). Una definición un tanto ambigua que levantó las expectativas de muchos, viendo en ella una vía para cuantificar el impacto social de los trabajos académicos de manera rápida y sencilla (Bornmann, 2014, Haustein, 2016). Asimismo, despertó un gran interés comercial, con grandes editoriales comprando y comercializando productos y proveedores de altmétricas. Entre ellos, los más reseñables son Altmetric.com, comercializado por Digital Science, una empresa perteneciente al grupo Springer Nature; y PlumX, perteneciente a Elsevier. Esta rápida acogida por parte de los proveedores, hizo que la definición original se fuera desvirtuando con el tiempo (Rousseau y Ye, 2013) a favor de un mayor esfuerzo por capturar atención o impacto fuera del entorno académico. Así, los distintos proveedores han ido incluyendo en sus paquetes distintos indicadores que trascienden el ámbito digital como son las citas recibidas en informes, las menciones en medios de comunicación (Zahedi y Costas, 2018), citas en material docente (Torres-Salinas, Gorraiz y Robinson-Garcia, 2017) o presencia en catálogos de biblioteca entre otros (Torres-Salinas, Gumpenberger y Gorraiz, 2017) entre otros.

---

[1] Traducido al español por los autores. La cita original en inglés dice así: "is the study and use of scholarly impact measures based on activity in online tools and environments."

Sin embargo, la crítica no se hizo esperar (Sugimoto et al., 2017), y rápidamente salieron a la luz diversos análisis y estudios alertando sobre las importantes limitaciones que plantean las altmétricas cuando se utilizan con el objetivo de medir el impacto social de la ciencia (Haustein et al., 2016; Robinson-Garcia et al., 2017; Vainio & Holmberg, 2017; Zahedi, Costas y Wouters, 2014). Lejos de suponer un abandono del estudio de estos indicadores, estos estudios han abierto el ámbito para utilizar estas nuevas medidas de manera más creativa y ajustar las expectativas sobre lo que ofrecen las altmétricas y qué es exactamente lo que reflejan. En este sentido, las distintas propuestas actuales se circunscriben al análisis de tres ámbitos (Robinson-Garcia, 2019):

1. **El compromiso social de los investigadores.** Entendido como el reflejo en las redes sociales de una actitud proactiva por parte del investigador para conectar con agentes no académicos y difundir su trabajo científico en el entorno social. Un ejemplo de este tipo de aplicaciones es incorporar el análisis de redes sociales para caracterizar las comunidades de usuarios con las que interactúan los investigadores en las plataformas sociales (Robinson-Garcia, van Leeuwen y Ràfols, 2018).

2. **Identificación de temáticas socialmente relevantes.** Donde el interés radica en identificar los temas científicos que mayor discusión están generando en las distintas plataformas sociales. Este tipo de análisis pone de relevancia que el interés social por la ciencia viene determinado en gran parte, por la temática científica (Noyons, 2018), la localización geográfica (Costas et al., 2017) o el sector de la población (Robinson-Garcia, Arroyo-Machado y Torres-Salinas, 2019).

3. **Comunidades de atención**. Entendido como la caracterización del que consume la literatura científica. Estas aplicaciones se centran en el análisis del perfil de usuarios que comparten y mencionan trabajos científicos con el fin de identificar tipos de usuarios o comunidades (Diaz-Faes, Bowman y Costas, 2019; Joubert y Costas, 2019; Zahedi y Van Eck, 2014).

Los análisis altmétricos suponen una oportunidad ineludible para comprender mejor cómo, qué y quién consume la literatura científica que se produce. En el ámbito de las Ciencias Sociales además, tienen una significancia mayor por su doble vocación académica y de análisis de la realidad y el entorno social. El objetivo de este capítulo es ofrecer un análisis sintético y crítico de la atención mediática y social que ha recibido la producción científica de la sociología desde 2000 hasta 2018. En estos casi veinte años, se han producido más de 4,000 trabajos indexados en la Web of Science, generando casi 10,000 menciones en las diferentes plataformas sociales. A continuación ofreceremos una visión global sobre la atención social que la Sociología española a generado en estos años.

## 2. Material y métodos

Para este análisis hemos empleado el mismo set de publicaciones descrito y presentado el capítulo de Jiménez-Contreras, Repiso y Ruiz-Pérez en este mismo libro. Esto supone un total de 4,042 publicaciones indexadas en la base de datos Web of Science y publicadas entre 2000 y 2018. Para analizar la atención social generada, hemos utilizado el agregador de altmétricas Altmetric.com. Esta plataforma permite lanzar consultas de menciones en plataformas sociales y fuentes alternativas como son Twitter, Facebook, Wikipedia, citas en informes, medios de comunicación, patentes o programas docentes entre otros. La Tabla 1 incluye una definición de cada una de las fuentes cubiertas por Altmetric.com así como el peso que se le otorga a cada una de cara al cálculo del AAS.

**Tabla I**. *Indicadores altmétricos considerados incluidos en la plataforma Altmetric.com y a su vez considerados en el cálculo del Altmetric Attention Score*

| Fuente | Tipo | Peso | Actualización | Descripción |
|---|---|---|---|---|
| Blogs | Blog | 5 | Diaria | Menciones en un listado curado de 14.000 blogs |
| Policy documents | Informe | 3 | Diaria | Número de veces que un trabajo aparece en informes |
| Twitter | Medio social | 1 | Tiempo real | Número de tuits y retuits realizados a trabajos científicos. |
| Facebook | Medio social | 0,25 | Diaria | Número de menciones disponibles en páginas públicas |
| Reddit | Medio social | 0,25 | Diario | Menciones en las entradas originales de Reddit |
| YouTube | Multimedia | 0,25 | Diario | |
| Mainstream media | Noticias | 8 | Tiempo real | Menciones en las noticias de un listado de más 2.900 medios |
| Open Syllabus | Otros | 1 | Trimestral | menciones en los programas de estudios de 4.000 instituciones |
| Q&A | Otros | 0,25 | Diario | Menciones realizadas en hilos de Stack Exchange. |
| Patentes | Patentes | 3 | Mensual | referencias en patentes registradas en IFI CLAIMS. |
| PubPeer | Post-peer-review | 1 | - | Post-publication peer-review realizadas al trabajo en PubPeer. |
| Publons | Post-peer-review | 1 | - | Post-publication peer-review realizadas al trabajo en Publons. |
| F1000Prime | Recomendaciones | 1 | Diario | Veces que un trabajo es recomendado en F1000Prime. |
| Wikipedia | Wiki | 3 | Tiempo real | Menciones desde las entradas de Wikipedia (EN) |
| Otros: Pinterest, LinkedIn, Google+, Sin Weibo | Medio social | 0,25 \| 0,5 \| 1 \| 1 | Sin actualización | Número de menciones. Solo datos históricos |

Una limitación importante de este proveedor, es que sólo permite consultar las menciones de documentos que cuentan con identificador único (Robinson-Garcia et al., 2014). En nuestro caso, hemos empleado el Digital Object Identifier o DOI, el identificador de artículos de revista ampliamente expandido en el mundo de la publicación científica. No obstante, en nuestro set de datos, sólo 2,693 publicaciones cuentan con dicho identificador. De las restantes, observamos que el 49% provienen de revistas españolas y el 25% son tipos documentales no citables (e.g., reseñas de libros, resúmenes de congreso, biografías, ponencias de congreso, etc.). De aquellas con DOI, fueron localizadas 1,108 publicaciones en Altmetric.com y 920 de ellas contaban con al menos una mención en alguna de las fuentes que recoge la plataforma.

Paralelamente, hemos descargado para el periodo 2014-2018 en la suite bibliométrica de Web of Science InCites, la producción española en las tres categorías temáticas de Web of Science en las que más publican los sociólogos españoles después de Sociología. Estas son Demografía, Ciencias Políticas y Ciencias Sociales, Interdisciplinar. Hemos obtenido un total de 4,109 trabajos de los cuáles 1,162 han recibido al menos una mención según Altmetric.com. Esto mismo se ha hecho para la producción en Europa, Estados Unidos y América Latina en la categoría de Sociología para el mismo periodo. En este caso se trata de un total de 47,613 registros de Web of Science de los cuales 18,045 tienen menciones en Altmetric.com. Este proceso queda resumido en la Figura 1.

**Figura 1**. *Diagrama resumen del proceso para la recopilación, procesamiento y tratamiento de los registros bibliográficos y el cálculo de los indicadores alternativos*

Para la descripción de la producción científica se han empleado cuatro indicadores. El primero es el *Nº o % de documentos indexados en Altmetric.com* y que denota la cobertura de esta fuente. El segundo indicado es el Nº y % de documentos mencionados en Altmetric.com. En tercer lugar, como representante de la plataformas altmétricas se utiliza, por su mayor de grado de cobertura el *Nº de menciones recibidas en Twitter*. Finalmente, en cuarto lugar, a lo largo de todo el trabajo emplearemos el Altmetric Attention Score (AAS) un indicador agregado de todas las menciones que ofrece Altmetric. Se calcula para cada, en este trabajo lo emplearemos como filtro para identificar las publicaciones que más atención social reciben. Este indicador realiza una suma ponderada de las distintas menciones que un trabajo recibe, otorgando un mayor o menor peso a cada una de ellas en función de su importancia y limitaciones. De esta forma no todos los medios tienen el mismo valor.

El análisis estadístico y descriptivo así como la generación de las redes se han llevado a cabo empleando el lenguaje de programación R (versión 3.6.1). La identificación temáticase ha empleado utilizando el software Bibliometrix (Aria & Cucurullo, 2010) y para el análisis y visualización de redes se ha usado el software Gephi (versión 0.9.2).

# 3. Resultados

## 3.1. Datos generales

El total de artículos indexados en Altmetric.com es de 1,108, representando apenas un 26% del set de publicaciones completo. De este conjunto, 920 publicaciones cuentan con al menos una mención en alguna de las plataformas sociales recogidas en esta fuente. Sin embargo, la baja cobertura encontrada es dependiente de diferentes propiedades de los documentos (Tabla 2). El primer factor es el idioma, ya que hay una mayor presencia de trabajos publicados en inglés, para los que la cobertura se incrementa al 37%. Los trabajos publicados en español tan solo tienen una cobertura del 6,12%, ya que de los 1,241 trabajos en lengua española de nuestro set de publicaciones, sólo 76 fueron identificados en Altmetric.com. El segundo factor es la tipología documental. Los trabajos identificados como artículos tienen una mayor cobertura que el resto (35%).

**Tabla II.** *Cobertura de la Sociología española en la base de datos Altmetric.com considerando el idioma y la tipología documental.*

|  | Nº Documentos | Nº - % indexados en Altmetric.com |
|---|---|---|
| **(a) Según idioma** | | |
| Inglés | 2729 | 1030 - 37,74% |
| Español | 1241 | 76 - 6,12% |
| Francés | 30 | 2 - 6,67% |
| Gallego o Catalán | 22 | 0 - 0.00% |
| Otros idiomas europeos | 20 | 0 - 0.00% |
| **(b) Según tipología documental** | | |
| Artículo | 2655 | 933 - 35.14% |
| Capítulo de libro | 591 | 85 - 7.67% |
| Reseña de libro | 443 | 30 - 2.71% |
| Material editorial | 138 | 27 - 2.44% |
| Material editorial; Capítulo de libro | 106 | 0 - 0.00% |
| Otros | 109 | 6 - 0.02% |

El total de menciones recibidas por los trabajos de sociología es de 9,151. Esto excluye citas provenientes de la base de datos Dimensions así como Mendeley, cuya cobertura en Altmetric.com no es exhaustiva (Robinson-Garcia et al., 2014). Es importante señalar que si analizamos este conjunto de menciones el 90% de las mismas pertenecen a única plataforma, Twitter (Tabla 3). Esta red social acumula 8,284 menciones dirigidas a 804 artículos diferentes. Con porcentajes significativamente inferiores le siguen Facebook, menciones de noticias y blogs que solo acumulan respectivamente el 3.48%, el 2.37% y el 1.78%. Si observamos el periodo de atención social que cubre Altmetric.com, vemos que las primeras menciones identificadas datan

de comienzos del año 2012, con picos en Noviembre de 2013, septiembre de 2017 y Enero de 2018.

**Tabla III.** *Total de menciones recibidas en Altmetric.com por parte de los trabajos Sociología distribuidas entre las principales fuentes.*

| Plataforma | Menciones Totales | Nº Documentos mencionados | % Documentos mencionados | Media y sd | Mediana |
|---|---|---|---|---|---|
| Tweet | 8284 | 804 | 90.53% | 7,48 (± 22,87) | 2 |
| Facebook post | 318 | 194 | 3.48% | 0,29 (± 0,96) | 0 |
| News story | 217 | 54 | 2.37% | 0,2 (± 1,81) | 0 |
| Blog post | 163 | 112 | 1.78% | 0,15 (± 0,65) | 0 |
| Policy document | 76 | 59 | 0.83% | 0,07 (± 0,35) | 0 |
| Wikipedia page | 45 | 37 | 0.49% | 0,04 (± 0,25) | 0 |
| Resto de fuente (nº 6) | 48 | -- | -- | -- | -- |

### 3.2. Comparativa nacional e internacional

En esta sección ofrecemos una primera panorámica de los resultados obtenidos. Para ello, realizamos diferentes comparativas con el objeto de contextualizar y facilitar la interpretación de los resultados. Hacemos dos tipos de comparaciones (Tabla 4). La primera es de carácter disciplinar, donde comparamos los datos con tres disciplinas de las Ciencias Sociales a nivel nacional. La segunda comparativa es de carácter internacional, donde comparamos los datos de la Sociología española con los datos de la categoría Web of Science "*Sociology*" en tres ámbitos regionales diferentes (Estados Unidos, Unión Europea y Latinoamérica). En estos análisis, constreñimos el periodo cronológico a 2014-2018, considerando solo los trabajos con al menos una mención y empleando el Altmetric Attention Score como indicador principal.

En la comparativa disciplinar, observamos que la Sociología española presenta un perfil similar al de Ciencias Políticas y Ciencias Sociales (Interdisciplinar). La tasa de cobertura para el período se sitúan en estas tres áreas áreas entre el 32% y el 34% y solo Demografía se sitúa por encima (55%). Algo similar sucede cuando observamos el porcentaje de trabajos mencionados. El AAS promedio para la Sociología española es 7,95 por trabajo, por encima de las Ciencias Sociales (Interdisciplinar) pero por debajo de la Ciencia Política (8,49), si bien la mediana es de 3 en el caso de la Sociología y esta última disciplina.

Si comparamos con el contexto internacional, se observa que en la mayor parte de los indicadores son siempre favorables a Estados Unidos y la Unión Europea. Las tasas de cobertura en Altmetric.com en EEUU y la UE se acercan al 50%, muy lejos del 33% de la Sociología. Asimismo el impacto es muy superior en EEUU donde el promedio de AAS es de 14,15 frente al 7,95 de la Sociología, si bien mejora la mediana de la UE situada en 2. En el caso de Latinoamérica la Sociología se sitúa por encima en todos los indicadores. En este caso las tasas coberturas son muy similares y la diferencias se encuentran sobre todo en el promedio AAS. Los datos por tanto parecen consistentes y razonables, ya que sitúan a la Sociología española en un lugar intermedio, por encima de Latinoamérica, por debajo pero cercana a la unión europea pero todavía lejos de equiparar los indicadores a los de Estados Unidos.

**Tabla IV.** *Estadísticos del AAS de Altmetric.com comparados a nivel nacional con disciplinas y a nivel internacional con diferentes zonas geográficas*

|  | Nº Documentos | Indexados en Altmetric | % cobertura | % mencionados | Media y sd AAS | Mediana AAS |
|---|---|---|---|---|---|---|
| Sociología ES (**) | 4042 | 1108 | %27 | %22 | 7,24 (±23,56) | 2 |
| Sociología ES (**) 2014-2018 | 2204 | 738 | %33 | %29 | 7,95 (±22,89) | 3 |
| **(a) Comparativa nacional con categorías Web of Science** | | | | | | |
| Demography (WoS) | 348 | 192 | %55 | %42 | 10,7 (±64,48) | 2 |
| Political Science (WoS) | 1893 | 619 | %32 | %26 | 8,49 (±22,6) | 3 |
| SS, Interdisciplinary (WoS) | 1868 | 652 | %34 | %27 | 6,3 (±20,16) | 2 |
| **(b) Comparativa internacionales con la categoría Web of Science Sociology en:** | | | | | | |
| Estados Unidos | 21756 | 10238 | %47 | %41 | 14,15 (±60,54) | 3 |
| Europa | 24456 | 10891 | %44 | %35 | 9,19 (±27,85) | 2 |
| Latinoamérica | 1401 | 450 | %32 | %25 | 6,36 (±15,42) | 2 |

** Datos del conjunto de trabajos de los autores

### 3.3. Análisis de comunidades y temas

En esta sección centramos el análisis en la identificación y caracterización de usuarios comentando y mencionando trabajos españoles de Sociología, así como en la identificación de las temáticas que mayor atención social despiertan. La mayor parte del análisis se centra en la plataforma Twitter, ya que es la que mayor actividad genera de las distintas plataformas analizadas (Tabla 3). En la siguiente subsección nos centramos en el análisis de usuarios de esta plataforma. En líneas generales, observamos que la mayor parte de la actividad proviene de cuentas institucionales y editoriales, evidenciando una falta de actividad tanto de la propia comunidad de sociólogos como de otros agentes sociales. En la segunda subsección, identificamos los principales temas y trabajos discutidos. En este caso sí que evidencian los temas tratados un marcado carácter social.

#### 3.3.1. Comunidades de usuarios de Twitter

La mayor parte de menciones a los artículos científicos de Sociología provenían de referencias realizadas en tuits. En total han participado en las 8,284 menciones un total de 4,869 cuentas diferentes de Twitter, lo que evidencia una gran dispersión de usuarios. Es decir, no se observa una comunidad de individuos cohesionada que tuitee de manera activa sobre temas relacionados con la Sociología española.

La cuenta que más trabajos únicos ha referenciado es la de *@poblacion_csic* que ha emitido un total de 29 menciones. Junto a *@Demografia_CSIC* con 20 menciones, son las dos cuentas más activas con un perfil vinculado a una institución de investigación. De los perfiles más activos y más entregados a la difusión de los trabajos a través de Twitter encontramos sobre todo a las propias revistas científicas o grandes editores. Aquí destaca @*SAGEsociology* con 25 menciones junto a tres perfiles de revistas en el top 10 de usuarios de Twitter (Tabla 5). Estas revistas son: *Ethnic and Racial Studies*, *Wiley Politics* y *European Sociological Review*.

**Tabla V.** *Ranking top 10 de usuarios de Twitter que mayor número de menciones únicas han emitido a trabajos de Sociología españoles*

| Rango | Menciones únicas | Cuenta | Nombre | Seguidores | Tipo |
|---|---|---|---|---|---|
| 1 | 29 | poblacion_csic | Población CSIC | 3324 | Institucional |
| 2 | 25 | SAGEsociology | SAGE Sociology | 41339 | Revista / Editorial |
| 3 | 20 | madmakko | marco albertini | 1083 | Personal |
| 4 | 20 | Demografia_CSIC | Demografía (CSIC) | 6140 | Institucional |
| 5 | 19 | PopulationEU | Population Europe | 4133 | Institucional |
| 6 | 16 | Socio101 | RodBScherich | 1709 | Personal |
| 7 | 16 | ERSjournal | Ethnic and Racial Studies | 4232 | Revista / Editorial |
| 8 | 15 | WileyPolitics | Wiley Politics | 7233 | Revista / Editorial |
| 9 | 15 | ADRIANSYSNET | Adrian Sung | 1553 | Personal |
| 10 | 14 | ESR_news | ESR | 2079 | Revista / Editorial |

En la red se producen diversas singularidades. La primera es que los perfiles más activos difundiendo trabajos en español no se corresponden a autores o actores sociales españoles. En efecto, usuarios como @*madmakko*, @*Socio101* y @*ADRIANSYSNET* que corresponden a las cuentas individuales más activas, corresponden a profesores de universidades italiana, estadounidense y australiana respectivamente. En estos casos la difusión responde a un interés académico en el tema de investigación de los trabajos, no observando un interés personal por la difusión de trabajos propios.

No obstante, si consideramos las menciones totales (material complementario, tabla x) sí detectamos ciertas anomalías. Por ejemplo, la cuenta que realiza un mayor número menciones es la revista *The Sociological Review* a traves del usuario @*TheSocReview*. Esta cuenta menciona 148 veces 9 trabajos científicos de la propia revista. La cuenta @*flesherfominaya* realiza 114 menciones a 8 trabajos. Otro caso notorio es el de la cuenta de Josep LLuis Mico (@*JL_Mico*) que menciona un trabajo propio hasta 75 veces.

La Figura 2 muestra la red de usuarios de Twitter, relacionados a través de los trabajos que mencionan. Es decir, dos usuarios o nodos están conectados si tuitean un mismo trabajo de investigación. Como se puede observar, a pesar de las anomalías anteriormente mencionadas, sí que se observa cierta consistencia temática en los trabajos que los diferentes usuarios tuitean. El mayor grupo de usuarios centran su atención a los trabajos de Sociología relacionados con la Demografía y Estudios de la Familia (clúster azul), con una pequeña escisión de carácter más multidisciplinar (clúster naranja). Un segundo grupo, centra su atención a trabajos relacionados con Ciencias Políticas y Estudios Étnicos (clúster rosa). Una pequeña comunidad bastante cerrada es la centrada en temas relacionados con la salud (clúster rojo). Por su parte, los clústeres relacionados con Política medioambiental y educativa (clústeres negro y verde claro respectivamente) se entrecruzan, siendo dos comunidades no tan claramente delineadas como el resto. Aún así, observamos cierta dispersión con comunidades en muchos casos con muy pocos actores (e.g., el clúster de Sociología Interdisciplinar, color morado, está formado por tres cuentas, mientras que el naranja, cuenta con cinco usuarios).

**Figura 2.** *Agrupación de usuarios de Twitter a partir de co-menciones a trabajos científicos de sociólogos españoles.*

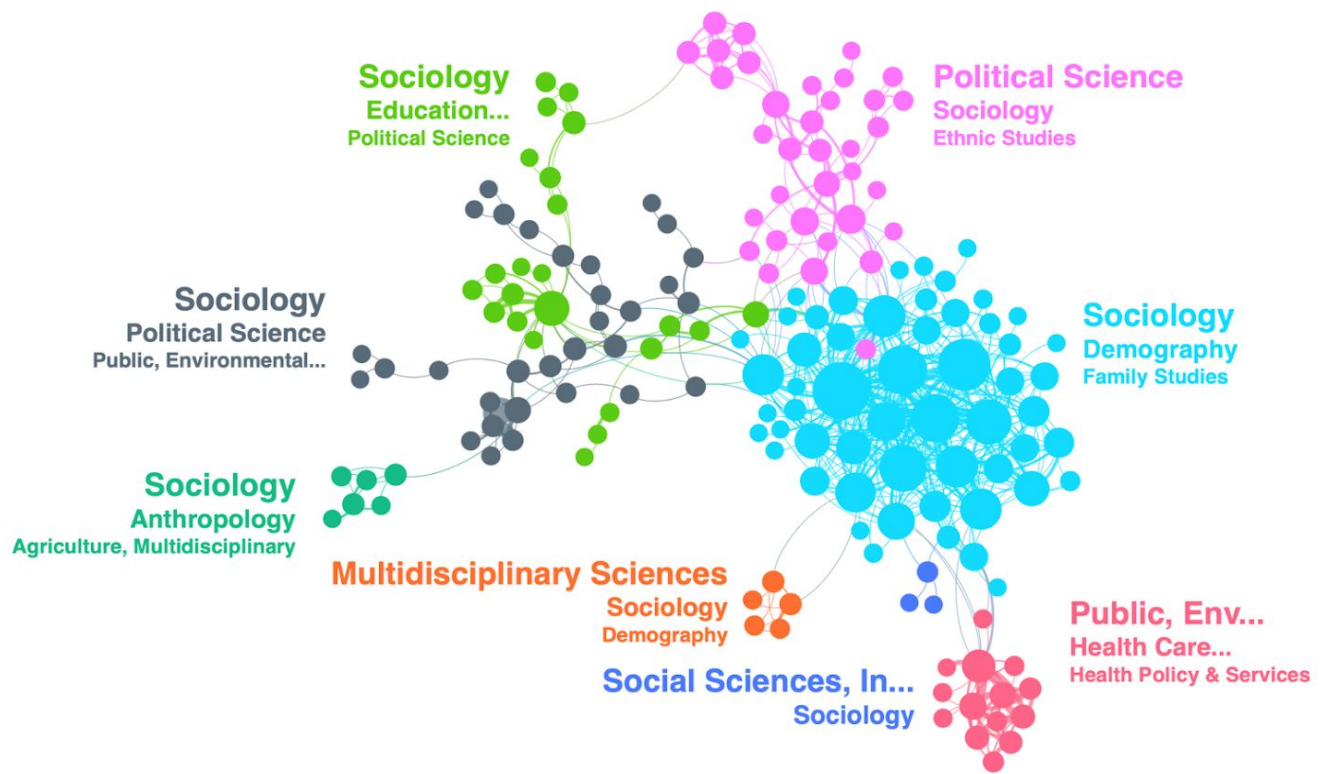

*Nota:* En color de los nodos corresponde al tema y categorías WoS predominantes. Comunidades detectadas mediante el algoritmo de Louvain (Q=0,591).

### 3.3.2. Análisis de temas y artículos destacados

Siguiendo con la identificación temática de los temas discutidos en Twitter, en la Figura 3 mostramos un mapa temático resultado de aplicar una técnica de Escalamiento Multidimensional o MDS. Incluimos sólo los trabajos con un AAS superior a 5. Es decir los temas está recibiendo una mayor atención, especialmente en Twitter. En líneas generales existe una gran claridad en los resultados. De los 13 trabajos identificados con un AAS > 50, 12 de ellos están integrados en alguno de los clústeres. Por tanto existe una clara coherencia entre la atención social que recibe un trabajo y el tema sobre el trata. Asimismo estas publicaciones científicas cuentas con unas características similares. Por un lado, están publicados en revistas de alto de impacto, preferente indexadas en el primer cuartil como por ejemplo American Sociological Review, Population and development review, Public Opinion Quarterly o European Sociological Review. En cuanto a la autoría, se imponen los trabajos publicados por universidades catalanas y en colaboración internacional con centros de prestigio como la universidades de Cambridge, Penn o Quebec.

**Figura 3.** *Principales conceptos de los trabajos con mayor Altmetric Attention Score (> 5). Técnica de visualización Multidimensional Scaling (MDS) a partir de las palabras clave de los trabajos*

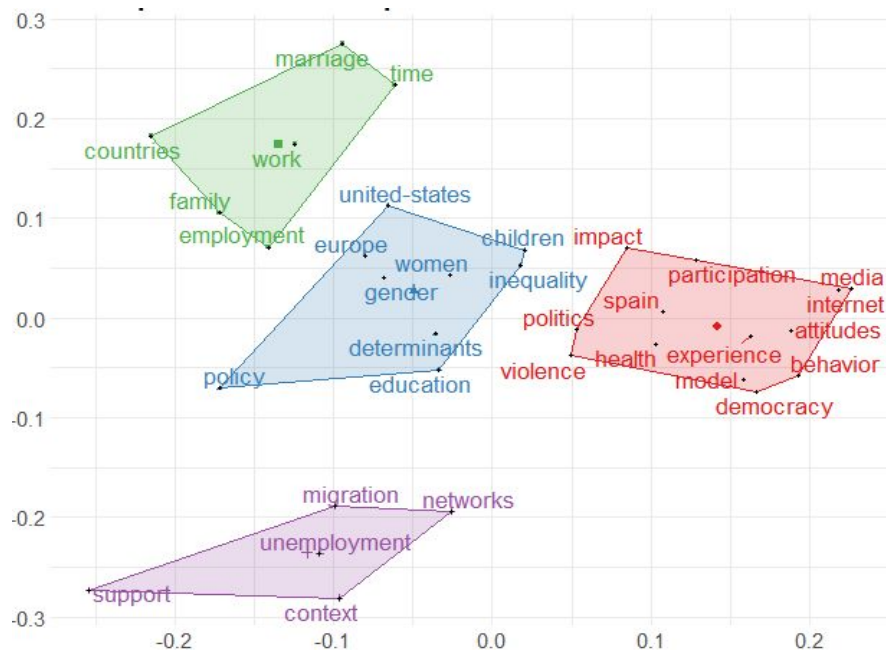

En la Figura 3 distinguimos cuatro ejes temáticos, tres de ellos bien cohesionados y un cuarto más heterogéneo. A continuación, de arriba a abajo, describimos cada uno de ellos:

- **Cluster verde**. Incluye como término central la palabra *work* (trabajo) asociada a téminos como *marriage* (matrimonio) o *family* (familia). Los artículos de este cluster están centrados en temas relacionados con la conciliación laboral, incluyendo la gestión del tiempo y estudios comparativos entre diferentes países. Este cluster, es el de mayor impacto de los cuatro identificados. Incluye los tres trabajos con mayor AAS. Así, el trabajo publicado por Sabino Kornrich cuenta con un AAS de 355, el mayor de los observados en nuestra colección. A continuación reseñamos los cuatro artículos más relevantes, incluyendo el único trabajo con un AAS > 50 publicado en una revista científica española (Reis: Revista Española de Investigaciones Sociológicas):

    - **AAS = 379**. Kornrich, S. (Instituto Juan March), Brines, J. (Univ. Washington), & Leupp, K. (Univ. Washington). (2013). *Egalitarianism, housework, and sexual frequency in marriage*. American Sociological Review, 78(1), 26-50.
    - **AAS = 137**. Martínez-Pastor, J. I. (UNED). (2017). *¿Importa el atractivo físico en el mercado matrimonial?*. Reis: Revista Española de Investigaciones Sociológicas, 91-111.

- **AAS = 139**. Flecha, R. (Univ. Barcelona), & Soler, M. (Univ. Barcelona) (2013). *Turning difficulties into possibilities: Engaging Roma families and students in school through dialogic learning*. Cambridge Journal of Education, 43(4), 451-465.
- **AAS=52**. Laplante, B. (Univ Quebec), Castro‑Martín, T. (CSIC), Cortina, C., & Martín‑García, T. (Pompeu Fabra) (2015). *Childbearing within marriage and consensual union in Latin America, 1980–2010*. Population and Development Review, 41(1), 85-108.

- **Cluster azul.** Este cluster trata temas muy cercanos al anterior. En este caso, se centra en estudios de género, así como desigualdades en la infancia y políticas educativas. Tiene un claro componente geográfico, con estudios localizados principalmente en Europa y Estados Unidos. Algunos de los términos centrales de este cluster son *gender*, *women*, *child* y *education*. Destacamos sobre todo un trabajo dedicado a políticas educativas para la prevención del terrorismo y otro sobre la brecha de género en la educación a nivel mundial y el fenómeno de la hipergamia. Las referencias de los trabajos mencionados son las siguientes:

    - **AAS=75**. **Citas: 11.** Aiello, E. (Univ Barcelona), Puigvert, L. (Univ Barcelona / Univ Cambridge), & Schubert, T. (Univ Rovira & Virgili). (2018). *Preventing violent radicalization of youth through dialogic evidence-based policies*. International sociology, 33(4), 435-453.
    - **AAS=64. Citas: 60.** Esteve, A. (Univ Autonoma Barcelona), Schwartz, C. R. (Univ Wisconsin), Van Bavel, J. (Univ Leuven), Permanyer, I. (Univ Autonoma Barcelona), Klesment, M. (Univ Leuven), & Garcia, J. (Univ Minnesota). (2016). *The end of hypergamy: Global trends and implications*. Population and development review, 42(4), 615.

- **Cluster rojo**. Es cluster de mayor tamaño por el número de conceptos que contiene y el más hetereogéneo semánticamente. En un primer subcluster identificamos temas de carácter político relativos a la participación ciudadana, la violencia o los modelos de democracia. Aparecen trabajos que versan sobre la utilización de redes neuronales para predecir la corrupción, el embargo de las encuestas electorales o el mapeo de ideologías políticas en los jóvenes. Existe asimismo un subcluster más relacionado con la utilización de internet y los nuevos medios de comunicación. En este bloque destaca especialmente un trabajo sobre el consumo de noticias digitales en España tras la introducción del impuesto por hiperenlacescon el fin de prevenir la piratería digital:

    - **AAS=112. Citas: 8.** López-Iturriaga, F. J. (Univ. Valladolid), & Sanz, I. P. (Higher School of Economics, Rusia). (2018). *Predicting public corruption with neural networks: An analysis of spanish provinces*. Social Indicators Research, 140(3), 975-998.
    - **AAS=96**. **Citas: 4.** Lago, I. (Univ Pompeu Fabra), Guinjoan, M. (Univ Autonoma Barcelona), & Bermúdez, S. (Univ Pompeu Fabra). (2015). *Regulating Disinformation: Poll Embargo and Electoral Coordination*. Public Opinion Quarterly, 79(4), 932-951
    - **AAS=71**. **Citas: 14.** Pollock, G. (FP7 Project MYWEB, Barcelona), Brock, T. (Manchester Metropolitan Univ), & Ellison, M. (Manchester Metropolitan Univ)

  (2015). *Populism, ideology and contradiction: mapping young people's political views*. The Sociological Review, 63, 141-166. (ASS = 71)
  - **AAS=60**. **Citas: 81.** Bacigalupe, A. (Euskal Herriko Unibertsitatea), & Escolar-Pujolar, A. (independiente). (2014). *The impact of economic crises on social inequalities in health: what do we know so far?*. International journal for equity in health, 13(1), 52.
  - **AAS=55**. **Citas: 8** Majó-Vázquez, S. (Univ Oxford), Cardenal, A.S. (Univ Oberta Catalunya), & González-Bailón, S. (Univ Penn) (2017). *Digital news consumption and copyright intervention: Evidence from Spain before and after the 2015 "link tax"*. Journal of Computer-Mediated Communication, 22(5), 284-301.

- **Cluster morado.** Aparece sensiblemente aislado de los descritos anteriormente e integrado por cinco conceptos relacionados con el desempleo. Más concretamente el cluster versa sobre la relación entre el empleo y la inmigración, haciendo hincapié en el contexto de los inmigrantes y sus redes de apoyo. Un claro ejemplo de los artículos de este cluster es el siguiente trabajo, que cuenta con el mayor AAS en esta temática:

  - **AAS=53**. **Citas: 2.** Polavieja, J. G. (Univ Carlos III), Fernández-Reino, M., & Ramos, M. (Univ Carlos III). (2018). *Are migrants selected on motivational orientations? Selectivity patterns amongst international migrants in Europe*. European Sociological Review, 34(5), 570-588.

Finalmente, resaltamos otros trabajos singulares de la colección atendiendo ahora a su impacto en una fuente social concreta. Destaca por su difusión en medios de noticias (10 menciones) el trabajo de Núñez (2019) sobre sacerdotes que abandonan la iglesia, reseñado en medios como *The Huffington Post* o US News. El más citado en informes de organismos nacionales y supranacionales (policy documents) es el de Kohler (2002) sobre fertilidad en Europa en los '90, que ha recibido seis menciones en informes publicados por instituciones como *Analysis & Policy Observatory* (2019, 2005), *United States Census Bureau* (2016, 2008) o *International Monetary Fund* (2007). Son pocos los trabajos mencionados en Wikipedia, pero el artículo más destacado es el trabajo de Saris y Revilla (2016) sobre errores en encuestas y que ha sido citado en cuatro entradas de la wikipedia en su versión inglesa (Observational error, Willem Saris, Survey - human research y Questionnaire). Las referencias completas de los tres trabajos referenciados son las siguientes:

- **AAS=90. Citas: 5.** Nuñez, F. (Univ Barcelona). (2010). *Leaving the Institution. Secularized Priests*. Social Compass, 57(2), 268-284.
- **AAS=18**. **Citas: 1563.** Kohler, H. P. (Univ Pennsylvania), Billari, F. C. (Univ Mila), & Ortega, J. A. (Univ Autónoma Madrid) (2002). *The emergence of lowest‑low fertility in Europe during the 1990s*. Population and development review, 28(4), 641-680.
- **AAS=3**. **Citas: 35.** Saris, W. E. (Univ Pompeu Fabra), & Revilla, M. (Univ Pompeu Fabra). (2016). *Correction for measurement errors in survey research: necessary and possible*. Social Indicators Research, 127(3), 1005-1020.

# 4. Discusión y recomendaciones

## 4.1. Discusión de los resultados

En este capítulo se ha afrontado el reto de estudiar cuantitativamente una disciplina con peso significativo en las ciencias sociales españolas bajo la lupa de las altmétricas; éstas medidas pertenecen a un conjunto de indicadores de nuevo cuño sobre los que existen grandes expectativas, sobre todo por parte de la UE, para resolver el problema de la medición del impacto social (*societal impact*) (European Union, 2017), pero que aún cuentan con serias limitaciones. Los resultados presentados en este ensayo ilustran claramente éstas dos vertientes, por un lado la evidente utilidad de las altmetricas para capturar la "atención" o "influencia" de los trabajos en el mundo digital pero, por otro lado, también se han mostrado sus limitaciones metodológicas como, por ejemplo, la dependencia de los DOI, las bajas coberturas de algunas fuentes o los problemas de manipulación. Por ello los resultados alcanzados debe interpretase como una aproximación experimental e ilustrativa de los futuros contextos evaluativos a los que, sin duda, alguna nos veremos enfrentados en el futuro. A continuación realizamos una serie de observaciones que nos permitan contextualizarlo dichos resultados.

Desde el punto de vista de las altmétricas la sociología española comparte características similares a las obtenidas para otras disciplinas y otros conjuntos de datos. Uno de los fenómenos habituales es la concentración que se produce del total de menciones que recibe un conjunto de documentos en una plataforma concreta; así en el caso de la sociología se ha verificado como el 90.53% del total de las menciones que reciben los artículos pertenecen *Twitter*. Este patrón, si bien más acusado en la sociología, se produce por cuando estudiamos la producción científica de las universidades españolas; en un estudio que consideraba un total 125.824 artículos Web of Science publicados por el sistema universitario español fue 82% (Torres-Salinas et al., 2018). Por otro lado, si continuamos la comparativa con el trabajo de Torres-Salinas, en el mismo se señala que la tasa de cobertura de las publicaciones en la base de datos de Altmetric.com, en el caso de las universidades, es del 42% frente al 33% de la Sociología. En líneas generales la tasa de cobertura de la sociología se equipara a áreas similares en España como Ciencias Políticas (32%) o Ciencias Sociales, Interdisciplinar (33%), pero no llega a homologarse con las tasas de cobertura de Europa o EEUU que se sitúan en el 44% y el 47%, un valor más cercano al de las universidades españolas.

Los estudios más recientes analizando el SSCI y el A&HCI para un período similar (2013–2017) revelan que el 55.32% de las publicaciones con DOI están en Altmetric (Yang y Zheng, 2019). Las comparativas internacionales con otras regiones muestran el amplio margen de mejora que existe en relación a la cobertura. Una de las explicaciones de tasa de cobertura radica en que solo un 60% de los trabajos analizados cuentan con DOI un valor claramente inferior al 90% identificado estudios previos Gorraiz et al. (2016). Si bien los primeros análisis de cobertura del DOI apuntaban que en Ciencias Sociales la tasa de trabajos con DOI es sensiblemente menor a la de otras disciplinas (Haustein, Bowman y Costas, 2015) el estudio, ya referenciado de Yang y Zheng (2019) para un total de 1076154 publicaciones nos señala que el 90% de los artículos de ciencias sociales contaban con DOI.

También se han comprobado diferencias estadísticas en los valores del *Altmetric Attention Score;* las diferencias hay que buscarlas en la concentración de la producción científica de la sociología española en revistas científicas nacionales. Los resultados incluidos en el Material Complementario (MC Tabla I) y el apartado 3.3.2. apuntan a que los trabajos publicados en revistas internacionales publicadas en inglés suelen tener un mejor rendimiento en los indicadores almetrics. Por ejemplo la tasa de indexación es de los artículos en inglés del 37% mientras que en español del 6%; por ello la excesiva publicación en revistas nacionales y en español nos lleva a tener un perfil más cercano en sus valores de cobertura y AAS al contexto latinoamericano. Unas de las cuestiones importantes a analizar es si este tipo de métricos medidas de una plataforma *como Altmetric.com* son adecuadas para analizar una disciplina de las ciencias sociales a nivel nacional o existe algún tipo de sesgo anglosajón

Los datos demuestran que la Sociología se encuentra a una enorme distancia de Estados Unidos. Sin embargo, al igual que ocurría años atrás con la Web of Science, estas diferencias pueden estar provocadas por la selección de las plataformas y fuentes a partir de las cuales se calculan los indicadores. Por ejemplo *Altmetric.com* para recopilar las menciones de *Wikipedia* solo emplea la versión inglesa obviando todas las citas de la versión española. Igual ocurre en la selección de medios para las menciones incluidas en los indicadores News o Blogs Mentions que está claramente sesgados al contexto estadounidense. Asimismo, nos encontramos se utilizan plataformas como *Reddit* que claramente están circunscritas a Estados Unidos y asimismo también hay que reseñar la utilización de fuentes muy centrada en una disciplina como ocurre con la plataforma *F1000* que solo afecta a las Ciencias de la Salud. Esta situación no invalida los datos pero si los contextualiza ya que si bien existen muchas tipologías de altmetricas realmente la aplicación práctica de algunas de ellas, bien sea por el sesgo geopolítico o disciplinar, es bastante limitada.

Sin embargo debemos buscar explicaciones internas relacionadas con las dinámicas internas de la propia disciplina y el uso que realizan de las redes sociales y otras plataformas para la difusión de trabajos científicos. Una de las cuestiones reveladas por el trabajo es la poca cohesión, en términos de comunidades, que existe en la promoción científica de las publicaciones. Un indicador que nos revela esta situación es la enorme dispersión de las menciones en *Twitter* a través de 4869 cuentas o usuarios diferentes siendo las más relevantes extranjeras, bien sean éstas gestionadas por editoriales o por investigadores. Solo el CSIC a través de dos cuentas parece difundir trabajos de interés a sus comunidades. Asimismo se han detectado comportamientos de cuentas dirigidos a aumentar el número de tweet o menciones, lo que nos ha permitido cuestionarse la relevación de algunos de los identificados. Este tema ha sido discutido y evidenciado en otros trabajos (Robinson-García et al., 2017) que ya señalaban la importancia de contextualizar bien las métricas obtenidas en *Twitter*; la Sociología tampoco escapa a las problemas éticos y desviaciones de las altmetricas.

En relación a los temas y áreas de interés hemos podido establecer, a pesar de todo, que existen comunidades en *Twitter* muy vinculadas a la difusión de trabajos indexados en categoría WoS como *Political Science* y *Sociology/Demography* y *Public and Enviromental Health*. Este trabajo también ha permitido obtener un perfil de los artículos más difundidos y que alcanzan un mayor *Altmetric Attention Score* estableciéndose cuatro bloques de temas, especialmente atractivos para las plataformas sociales, que podemos denominar abreviadamente como Familia, Género, Política e Inmigración. Estos bloques representan hasta qué punto los intereses sociales y los científicos convergen. Considerando estos resultados había que preguntarse hasta qué punto pueden ayudar a definir este tipo de métricas las agendas de investigación del futuro. A fin de dilucidar estas cuestiones los sociólogos no deberían permanecer al margen de las propuestas altmétricas, y no solo deberían conocer los indicadores que se introducen en este trabajo sino que además deberían favorecer un contexto favorable a la medición. Por esta última razón, para cerrar este capítulo, hemos creído conveniente realizar una serie de recomendaciones orientadas a favorecer y mejorar la recopilación de este tipo de métricas.

## 4.2. Recomendaciones y buenas prácticas

A pesar del carácter exploratorio de este trabajo, hemos considerado una serie de puntos que consideramos de especial relevancia y que a continuación presentamos a modo de recomendaciones para maximizar los esfuerzos de difusión de la labor investigadora de la comunidad sociológica española.

1. **Las revistas españolas en el ámbito de la sociología deberían incluir de manera consistente y global, identificadores únicos de sus trabajos.** Prácticamente la mitad de los trabajos de sociólogos españoles carece de identificadores únicos que permitan rastrear y monitorizar la atención social recibida. Gran parte de estos trabajos provienen de revistas españolas, donde la asignación de DOIs no ha alcanzado el grado de integración que se observa en el ámbito internacional. Es necesario exigir este tipo de identificadores ya que no sólo facilitan la monitorización de trabajos, sino que también facilitan su difusión y visibilidad.

2. **La comunidad sociológica debería adoptar una actitud más proactiva de la difusión de su actividad investigadora en medios sociales.** Derivado de nuestro análisis de usuarios en Twitter, observamos una ausencia llamativa de sociólogos españoles discutiendo en las redes sus propios trabajos de investigación. Aunque la auto-referenciación excesiva nunca es deseable, sí que es necesario y recomendable que los investigadores sean más proactivos en la difusión de sus propios trabajos. E incluso, cuando no se trata de la difusión de trabajos propios, participen de la discusión global que se produce en estas redes sobre trabajos de su propio ámbito. Esto no sólo resulta enriquecedor en el plano personal y profesional, sino que otorga una visibilidad y una voz propia que sirva como referencia a actores sociales interesados en estas temáticas.

3. **Para elaborar una estrategia de difusión de resultados de investigación, se recomienda consistencia en la referenciación de trabajos así como un mayor grado de involucración.** Unido a los dos puntos anteriores tiene que ver cómo se hace dicha difusión. En este trabajo hemos utilizado nuevas métricas llamadas altmétricas para monitorizar la atención social que reciben los trabajos de los sociólogos españoles. Es cierto que estas métricas no están exentas de limitaciones, siendo una de ellas el gran número de menciones que no son capaces de capturar. Pero estas herramientas se están convirtiendo en un estándar cada vez más expandido. Por lo que si los investigadores sociólogos quieren hacer uso de plataformas sociales para difundir sus trabajos y luego poder mostrar dicha atención social a gestores universitarios y evaluadores, deberán ser rigurosos en el uso de DOIs y enlaces directos a los trabajos publicados, para asegurar que dichas menciones sean capturadas por los indicadores altmétricos.

4. **Las instituciones y revistas deben adoptar políticas articuladas con los propios investigadores de difusión.** En nuestro análisis observamos que gran parte de la actividad que se genera en torno a la Sociología española proviene de revistas e instituciones. No obstante, esta actitud proactiva de difusión parece caer en saco roto viendo la ausencia de investigadores y otro actores sociales (asociaciones, sociedades, instituciones no académicas). Algo resulta llamativo al ver que las temáticas tratadas tienen un eminente carácter social. Las políticas de difusión de carácter institucional o editorial deben venir articuladas y contar con el apoyo de la comunidad investigadora. En caso contrario se tratará de un trabajo inútil que no encontrará respuesta en la sociedad.

# 5. Referencias bibliográficas

# Material complementario

**MC Tabla I**. *Revistas científicas con mayor Altmetric Attention Score acumulado*

|    | Revista | AAS total | Media | Desviación estándar | Mediana | Nº artículos |
|----|---------|-----------|-------|---------------------|---------|--------------|
| 1  | AMERICAN JOURNAL OF SOCIOLOGY | 524 | 87.33 | 196.84 | 9.0 | 6 |
| 2  | AMERICAN SOCIOLOGICAL REVIEW | 485 | 97.00 | 159.59 | 45.0 | 5 |
| 3  | SOCIAL INDICATORS RESEARCH | 419 | 4.51 | 12.61 | 1.0 | 93 |
| 4  | EUROPEAN SOCIOLOGICAL REVIEW | 367 | 15.96 | 23.44 | 6.0 | 23 |
| 5  | SOCIOLOGICAL REVIEW | 364 | 30.33 | 45.50 | 9.0 | 12 |
| 6  | REIS | 264 | 12.57 | 29.89 | 2.0 | 21 |
| 7  | SOCIAL NETWORKS | 214 | 9.73 | 14.15 | 2.5 | 22 |
| 8  | BMJ-BRITISH MEDICAL JOURNAL | 203 | 203.00 | NA | 203.0 | 1 |
| 9  | POPULATION AND DEVELOPMENT REVIEW | 193 | 14.85 | 19.63 | 6.0 | 13 |
| 10 | REVISTA INTERNACIONAL DE SOCIOLOGIA | 163 | 4.08 | 8.45 | 1.0 | 40 |

**MC Figura 1**. *Evolución temporal del número de menciones en Altmetric. com de la producción científica española de sociología*

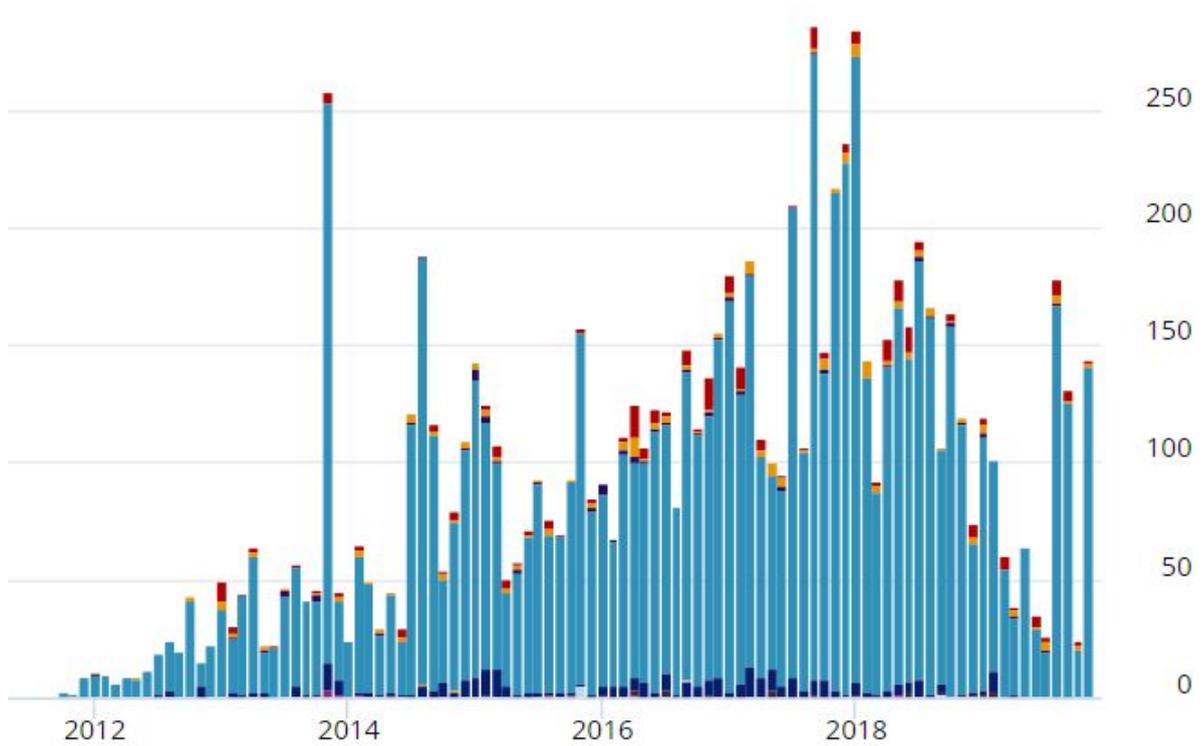

\* Figura extraída directamente de Altmetric.com.
En azul claro aparece representado el número de menciones en Twitter

**MC Tabla II**. *Cuentas de Twitter con mayor número de menciones totales y trabajos mencionados*

| Menciones totales | Trabajos mencionados | Cuenta | Nombre | Seguidores |
|---|---|---|---|---|
| 148 | 9 | TheSocReview | The Sociological Review | 52967 |
| 114 | 8 | flesherfominaya | C. Flesher Fominaya | 436 |
| 75 | 1 | JL_Mico | JL_Micó | 26636 |
| 43 | 25 | SAGEsociology | SAGE Sociology | 41339 |
| 39 | 29 | poblacion_csic | Población CSIC | 3324 |
| 36 | 14 | ESR_news | ESR | 2079 |
| 35 | 3 | IJSRM | IJ of SR Methodology | 2079 |
| 28 | 20 | Demografia_CSIC | Demografía (CSIC) | 6140 |
| 28 | 1 | HumansAnalytics | Humans Of Analytics | 3248 |
| 27 | 20 | madmakko | marco albertini | 1083 |
| 27 | 1 | itknowingness | IT Knowingness | 16687 |
| 24 | 19 | PopulationEU | Population Europe | 4133 |
| 21 | 1 | ICalzada | ICalzada | 3369 |
| 21 | 7 | SimulPast | SimulPast | 740 |
| 20 | 16 | ERSjournal | Ethnic and Racial Studies | 4232 |
| 20 | 3 | IS_sociology | InternationalSocio | 2036 |
| 20 | 11 | hcebolla | Hector Cebolla B. | 1333 |
| 19 | 4 | brianrahmer | Brian Rahmer | 8626 |
| 19 | 15 | ADRIANSYSNET | Adrian Sung | 1553 |
| 18 | 6 | isa_sociology | ISA | 25291 |
| 18 | 7 | CurrentSociolog | Current Sociology | 4263 |
| 18 | 5 | boscoaa | Anna Bosco | 1091 |
| 18 | 12 | aarcarons | Albert F. Arcarons | 1260 |
| 17 | 2 | acorsin | Alberto Corsín Jiménez | 3173 |
| 17 | 15 | WileyPolitics | Wiley Politics | 7233 |
| 17 | 7 | SAGEGender | SAGE Gender Studies | 1737 |